\documentclass[]{aa}
\usepackage{txfonts}
\usepackage{natbib}
\usepackage{graphicx}
\usepackage{psfig}

\def\XMM{{\em XMM--Newton}}
\def\Einstein{{\em Einstein}}
\def\SAX{{\em BeppoSAX}}
\def\ROSAT{{\em ROSAT}}
\def\RXTE{{\em RXTE}}
\def\Chandra{{\em Chandra}}
\def\Swift{{\em Swift}}
\def\OGLE{{\em OGLE}}
\def\ASCA{{\em ASCA}}

\def\pn{{\em pn}}

\def\pn{{\em pn}}

\def\sxp{SXP59.0}
\def\IGR{IGR J01572-7259}
\def\rxj{RX J0059.2-7138}
\def\smc{SMC X-2}
\def\approxgt{\mathrel{\hbox{\rlap{\lower.55ex \hbox {$\sim$}}
        \kern-.3em \raise.4ex \hbox{$>$}}}}
\def\approxlt{\mathrel{\hbox{\rlap{\lower.55ex \hbox {$\sim$}}
        \kern-.3em \raise.4ex \hbox{$<$}}}}
\def\ps {$P_{\rm spin}$}
\def\po {$P_{\rm orbit}$}
\def\pdot {$\dot P_{\rm spin}$}

\def\lx {$L_{\rm X}$}
\def\fx {$f_{\rm X}$}
\def\flux {\mbox{erg cm$^{-2}$ s$^{-1}$}}
\def\lum {\mbox{erg s$^{-1}$}}
\def\nh{$N_{\rm H}$}
\def\ltsima{$\; \buildrel < \over \sim \;$}
\def\lsim{\lower.5ex\hbox{\ltsima}}
\def\gtsima{$\; \buildrel > \over \sim \;$}
\def\gsim{\lower.5ex\hbox{\gtsima}}
\def\msole{~M_{\odot}}

\def\countsec{\hbox{counts s$^{-1}$}}
\def\fph {ph cm$^{-2}$ s$^{-1}$}
\def\hcm {\hbox {\ifmmode $ atom cm$^{-2}\else atom cm$^{-2}$\fi}}

\def\chisqnu {$\chi^{2}_{\nu}$}

\def \apjl {ApJL}
\def \apjs {ApJS}
\def \aap {A\&A}

\newcommand{\be}{\begin{equation}}

\newcommand{\ee}{\end{equation}}

\begin{document}
\title{Spectral analysis of \sxp\ during its 2017 outburst and properties of the soft excess in X-ray binary pulsars \thanks{Based on observations obtained with XMM-Newton, an ESA science mission with instruments and contributions directly funded by ESA Member States and NASA}}
\author{N.~La Palombara\inst{1}, P.~Esposito\inst{1,2}, F.~Pintore\inst{1}, L.~Sidoli\inst{1}, S. Mereghetti\inst{1}, and A.~Tiengo\inst{1,3,4}}

\institute{INAF, Istituto di Astrofisica Spaziale e Fisica Cosmica, via E.\ Bassini 15, I-20133 Milano, Italy
\and Anton Pannekoek Institute for Astronomy, University of Amsterdam, Postbus 94249, NL-1090-GE Amsterdam, The Netherlands
\and Scuola Universitaria Superiore IUSS Pavia, piazza della Vittoria 15, I-27100 Pavia, Italy
\and INFN, Sezione di Pavia, via A. Bassi 6, I-27100 Pavia, Italy}


\offprints{N. La Palombara, nicola.lapalombara@inaf.it}

\date{Received / Accepted}

\authorrunning{N. La Palombara et al.}

\titlerunning{Spectral analysis of \sxp}

\abstract{
We report the results provided by the \XMM\ observation of the X-ray binary pulsar SXP59.0 during its most recent outburst in April 2017. The source was detected at $f_{\rm X}$(0.2--12 keV) = 8$\times 10^{-11}$ \flux, one of its highest flux levels reported to date. The measured pulse period was \ps\ = 58.949(1) s, very similar to the periods measured in most of the previous observations. The pulsed emission was clearly detected over the whole energy range between 0.2 and 12 keV, but the pulse profile is energy dependent and the pulsed fraction increases as the energy increases. Although the time-averaged EPIC spectrum is dominated by a power-law component (with photon index $\Gamma = 0.76 \pm 0.01$), the data show an evident soft excess, which can be described with the sum of a black-body and a hot thermal plasma component (with temperatures $kT_{\rm BB} = 171^{+11}_{-14}$ eV and $kT_{\rm APEC} = 1.09^{+0.16}_{-0.09}$ keV, respectively). Moreover, the EPIC and RGS spectra show narrow emission lines due to N, O, Ne, Mg, and Fe. The phase-resolved spectral analysis of the EPIC data shows that the flux of the black-body component varies with the pulse phase, while the plasma component is almost constant. We show that the black-body component can be attributed to the reprocessing of the primary emission by the optically thick material at the inner edge of the accretion disc, while the hot plasma component is due to a diffuse gas far from the accretion region and the narrow emission lines of the RGS spectrum are most probably due to photoionized matter around the accreting source.

\keywords{accretion, accretion discs - stars: neutron - X-rays: binaries - pulsars: individual: \sxp\ - X-rays: individual (2E 0053.2--7242, XTE J0055--724, SAX J0054.9--7226, RX J0054.9--7226) - stars: individual: [MA93] 810}}


\maketitle

        \section{Introduction\label{intro}}

The X-ray pulsar \sxp\ was discovered with the \textit{Rossi X-ray Timing Explorer} (\RXTE) in January 1998, during an observation in the direction of the Small Magellanic Cloud (SMC). The source (designated as XTE J0055-724) was detected at a flux level \fx\ $\simeq 6 \times 10^{-11}$ \flux\ in the energy range 2--10 keV, and displayed a periodic modulation with \ps\ = 59.0 $\pm$ 0.2 s \citep{Marshall+98}. A follow-up observation with \SAX\ led to the detection of the bright source SAX J0054.9-7226 within the 10 arcmin error circle of XTE J0055-724, at a flux level \fx\ $\simeq 1.9 \times 10^{-11}$ \flux\ between 2 and 10 keV. The measurement of a strong pulsed emission with \ps\ = 58.969 $\pm$ 0.001 s confirmed the source association with XTE J0055-724 \citep{Santangelo+98}. The uncertainty on the position was about 40 arcsec, thus including the \ROSAT\ and \Einstein\ X-ray sources RX J0054.9-7226 (also known as 1WGA J0054.9-7226) and 2E 0053.2-7242. The latter was known to be variable on timescales from months to years and was proposed as a candidate High Mass X-ray Binary (HMXB) by \citet[source \#9]{Bruhweiler+87}, by \citet[source \#35]{WangWu92}, and by \citet{White+94}. The analysis of 13 archival \ROSAT\ PSPC and HRI observations revealed the former source three times between 1991 and 1996, at a luminosity level between $\simeq 8 \times 10^{34}$ and $4 \times10^{35}$ \lum\ (in the energy range 0.1--2.4 keV). Moreover, in the 1991 observation a pulse period \ps\ = 59.072 $\pm$ 0.003 s was measured. This result confirmed the association of the \RXTE\ and \SAX\ sources with RX J0054.9-7226 \citep{Israel+98}, which can very likely be identified with 2E 0053.2-7242.

Between 1998 and 1999 \RXTE\ detected four outbursts from \sxp, each with a duration of $\simeq$ 40 d. Based on the time spacing of these outburst, \citet{Laycock+05} derived an orbital period \po\ = 123 $\pm$ 1 d. Later on, this result was confirmed by \citet{Galache+08}, who extracted the orbit ephemeris and derived a period \po\ = 122.10 $\pm$ 0.38 d from five source outbursts observed with \RXTE\ between 2002 and 2004. During  the periods of activity in 1998--1999 and in 2002--2004 the source showed a short-term spin-up and reached a 3--10 keV luminosity \lx\ $\simeq 2 \times 10^{37}$ \lum. Moreover, although during the  $\sim$1100 days between the two groups of outbursts the source remained undetected, it experienced a long-term spin-up, with \pdot\ = -0.0227 $\pm$ 0.0006 s y$^{-1}$ \citep{Coe+10}. Finally, the source was detected again by \RXTE\ between the end of 2008 and the beginning of 2011, although at a lower luminosity level (\lx\ = 10$^{36-37}$ \lum). By taking into account all the \RXTE\ observations of this source over a time span of 13 years, \citet{Klus+14} measured an average pulse period and source luminosity of 58.859 $\pm$ 0.005 s and (8.4$\pm$0.2)$\times 10^{36}$ \lum, respectively. Throughout this time period \sxp\ experienced an average spin-up \pdot\ = -0.0206 $\pm$ 0.0005 s y$^{-1}$.

Three candidate counterparts were found within the 10 arcsec radius \ROSAT\ HRI error circle. All of them are early-type stars, but only one shows strong H$\alpha$ emission. Its radial velocity of 138 $\pm$ 27 km s$^{-1}$ is consistent with a location in the SMC. Moreover, it corresponds to object 810 of the catalogue of \citet{MeyssonnierAzzopardi93}, which includes the H$\alpha$ emission-line objects in the SMC. Therefore, this object was identified as the optical counterpart of \sxp\ \citep{Stevens+99} and, afterwards, it was classified as a B0e star \citep{Antoniou+09}. Moreover, it was extensively monitored with the \textit{MAssive Compact Halo Objects} (MACHO) and \textit{Optical Gravitational Lensing Experiment} (OGLE) projects. Based on the long-term MACHO and OGLE-II photometric data, \citet{SchmidtkeCowley05} proposed an orbital period \po\ = 60.2 $\pm$ 0.8 d, subsequently corrected to 62.15 $\pm$ 0.04 d with the addition of the OGLE-III data \citep{Rajoelimanana+11}. The optical period is half that of the X-ray, but no significant X-ray flux was detected half a phase from the X-ray maximum \citep{Galache+08}.

Since its discovery \sxp\ has been detected several times with the \textit{Advanced Satellite for Cosmology and Astrophysics} (\ASCA) \citep{Yokogawa+00}, the \textit{X-ray Multi-Mirror} (\XMM) mission \citep{Sasaki+03,Haberl+08,Sturm+13}, and the \Chandra\ telescope \citep{Laycock+10}. In their catalogue of the HMXBs in the SMC, \citet{HaberlSturm16} reported that the 0.2--10 keV flux of \sxp\ has varied between \fx$_{\rm min}$ = 7$\times 10^{-14}$ and \fx$_{\rm max}$ = 6$\times 10^{-11}$ \flux. Moreover, the catalogue of all the SMC observations performed with \RXTE, \XMM, and \Chandra, compiled by \citet{Yang+17}, shows that \sxp\ has been detected with \Chandra\ at even lower luminosity levels, down to \lx\ $\simeq 10^{34}$ \lum. The measured pulse periods imply a long-term spin-up of the pulsar, with \pdot\ = -0.016 $\pm$ 0.004 s y$^{-1}$.

On March 30, 2017 (MJD 57842), \Swift\ XRT detected an outburst from \sxp\ \citep{Kennea+17}. The measured count rate in the Photon Counting mode exposure was 0.9 $\pm$ 0.2 c s$^{-1}$. A follow-up observation, performed again with \Swift\ on April 7 (MJD 57850), found that this source had brightened since the previous observation, up to an XRT count rate of 2.1 $\pm$ 0.1 c s$^{-1}$. The detected spectrum was well fitted with an absorbed power-law (PL) model, with a photon index $\Gamma$ = 0.96 $\pm$ 0.10. The measured flux value (in the energy range 0.5--10 keV) was \fx\ = $8 \times 10^{-11}$ \flux, corresponding to \lx\ = $4 \times 10^{37}$ \lum\ for an SMC distance of 62 kpc \citep{Graczyk+14}. Therefore, we triggered our \XMM\ Target-of-Opportunity (ToO) programme for the observation of transient binary pulsars in the SMC. Here, we report on the results obtained with the follow-up \XMM\ observation of \sxp. 

         \section{Observation and data reduction}
         \label{data}

On April 14, 2017 (MJD 57857), two weeks after the beginning of the outburst, our target was observed with \XMM\ for a total exposure time of $\sim$ 14 ks. In Table~\ref{observation} we provide the set-up of the EPIC \pn\ \citep{Struder+01} and MOS \citep{Turner+01} focal-plane cameras and of the Reflection Grating Spectrometer (RGS, \citealt{denHerder+01}). All the collected events were processed with version 16 of the \XMM~{\em Science Analysis System}\footnote{https://xmm-tools.cosmos.esa.int/external/xmm\_user\_support/documentation/sas\_usg/USG/} (\textsc{sas}). We verified that the whole observation was free of soft-proton contamination, which could affect the data analysis. Therefore, we considered the full EPIC and RGS datasets for both timing and spectral analysis. The effective exposure times were $\simeq$ 10 ks for the \pn\ camera and $\simeq$ 14 ks for the MOS and the RGS instruments; for the EPIC cameras they take into account the dead time of 29 \% and 2.5 \% for \pn\ and MOS, respectively. In Table~\ref{observation} we summarize the \XMM\ observation.

In the case of the EPIC data, we selected mono- and bi-pixel events (with pattern between 0 and 4) for the \pn\ camera and from 1- to 4-pixel events (with pattern between 0 and 12) for the MOS cameras. From the spatial point of view, the events were selected from circular regions around the source position. In all cameras the CCD edges or dark columns limited the extraction radius, which was 35 arcsec for the \pn\ camera, 40 arcsec for the MOS1 camera, and 50 arcsec for the MOS2 camera. Although the source count rate (CR) was rather high (Table~\ref{observation}), we checked that the data of all the EPIC cameras were not affected by photon pile-up. To this end, we followed the same approach used by \citet{LaPalombara+18}: on the one hand, spectra with different pattern selections (only mono- or bi-pixel events), and on the other hand, spectra with or without the removal of the central part of the point spread function (PSF), where the possible pile-up is higher. In all cases the spectral analysis provided consistent results, thus showing that the pile-up was negligible. Therefore, for each camera the source events were selected  from the whole circular region. The corresponding background events were selected from circular regions which were offset from the target position and free of sources. In the \pn\ case it was on the same CCD of the source and, hence, it had a small radius of 35 arcsec. For the MOS cameras it was on a peripheral CCD and its radius was 200 and 250 arcsec for the MOS1 and MOS2, respectively.

\begin{table*}
\caption{Summary of the \XMM\ observation of \sxp\ (ID 0740071301).}\label{observation}
\vspace{-0.5 cm}
\begin{center}
\begin{tabular}{ccccccc} \\ \hline
Instrument      & Filter        & Mode                  & Time Resolution       & Net Exposure Time       & Extraction Radius     & Net Count Rate        \\
                        &                       &                               &                                       & (ks)                            & (arcsec)                      & (\countsec)           \\ \hline
\pn\            & Thin 1        & Small Window  & 5.7 ms                        & 10.0                            & 35                            & 9.49$\pm$0.03         \\
MOS1            & Thin 1        & Small Window  & 0.3 s                         & 13.8                            & 40                            & 2.64$\pm$0.01         \\
MOS2            & Thin 1        & Small Window  & 0.3 s                         & 14.0                            & 50                            & 3.02$\pm$0.01         \\
RGS1            & -             & Spectroscopy  & 4.8 s                         & 14.7                            & -                                     & 0.274$\pm$0.003 \\
RGS2            & -             & Spectroscopy  & 9.6 s                         & 14.6                            & -                                     & 0.090$\pm$0.002 \\ \hline
\end{tabular}
\end{center}
\end{table*}

         \section{Timing analysis}
         \label{timing}

\begin{figure}
\begin{center}
\includegraphics[width=5.75cm,angle=-90]{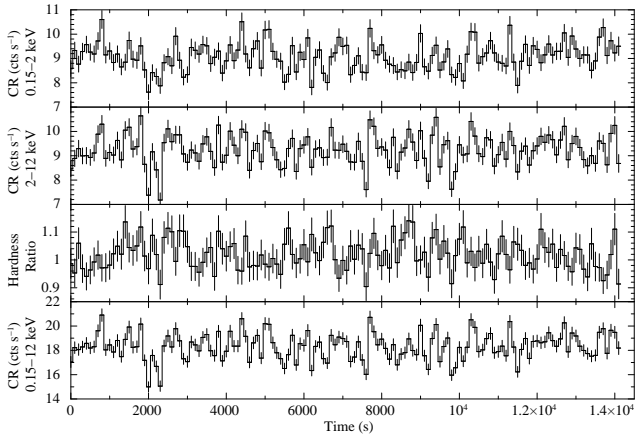}
\caption{Background-subtracted light curves of \sxp\ in the energy ranges 0.15--2, 2--12, and 0.15--12 keV, with a time binning of 100 s.}
\label{lc}
\end{center}
\vspace{-0.75 cm}
\end{figure}

For the timing analysis of the EPIC events, we converted the arrival times to the solar system barycentre  by using the \textsc{sas} tool \textsc{barycenter}. We defined three energy ranges (soft = 0.15--2 keV, hard = 2--12 keV, and total = 0.15--12 keV) and, for each range and each of the three cameras, we accumulated a light curve (with a time binning of 100 s). These curves were then corrected for the background and the extraction region by using the \textsc{sas} tool \textsc{epiclccorr}. In the total range the average CR was $\simeq$ 11.3, 3.6, and 3.5 cts s$^{-1}$, for the \pn, MOS1, and MOS2 cameras, respectively. For each of the three energy ranges, we calculated the cumulative light curve as the sum of the light curves of the individual cameras. We show the three curves in Fig.~\ref{lc}, together with the hardness ratio of the hard (H) to the soft (S) light curves (HR = H/S). In both the soft and hard energy ranges the average CR was $\simeq$ 9 cts s$^{-1}$. Although the source was variable over short timescales (with CR variations of up to $\sim$ 30 \% between consecutive time bins), its flux does not increase or decrease over the observation timescale. Moreover, no correlation is found between the HR and the CR: the HR shows some bin-to-bin variability, but without any dependence on the CR or clear trend on long timescales.

\begin{figure}
\begin{center}
\includegraphics[width=5.75cm,angle=-90]{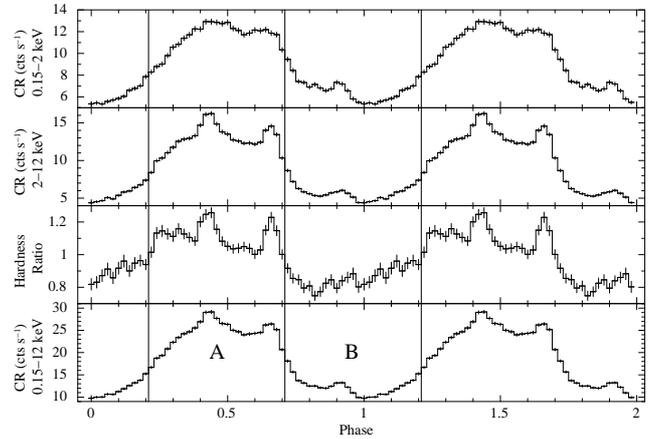}
\caption{Light curves of \sxp\ in the energy ranges 0.15--2, 2--12, and 0.15--12 keV, folded on the period $P$ = 58.949 s. The two phase intervals selected for the phase-resolved spectral analysis (section \ref{resolved_spectral_analysis}) are delimited by vertical lines.}
\label{flc2E}
\end{center}
\end{figure}

We merged the event datasets of the three instruments for the measurement of the pulse period of the source in order to increase the count statistics. Then, by applying a standard phase-fitting technique, we obtained a best-fitting period of $P$ = 58.949(1) s. The three folded light curves, and the corresponding HR, are shown in Fig.~\ref{flc2E}. The pulse profile is characterized by similar properties at all energies: it shows a broad maximum around phase $\Phi \simeq$ 0.5, a small peak at phase $\Phi \simeq$ 0.9, and a minimum around phase $\Phi \simeq$ 0. However, the figure shows some energy dependence since in the hard range two distinct narrow peaks arise above the broad maximum, at $\Phi \simeq$ 0.45 and 0.65. The HR is rather correlated with the CR and shows two narrow peaks coincident with those of the hard range. The average pulsed fraction, defined as PF = (CR$_{\rm max}$ - CR$_{\rm min}$)/(2$\times$CR$_{\rm average}$), is high and depends on energy, since it is $\simeq$ 42 \% for the soft range and $\simeq$ 64 \% for the hard range.

\begin{figure}
\begin{center}
\includegraphics[height=8.5cm,angle=-90]{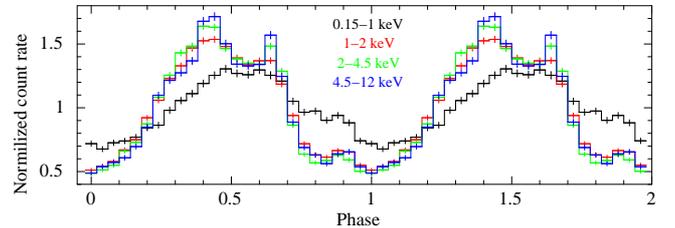}
\caption{Pulse profile of \sxp\ in the energy ranges 0.15--1, 1--2, 2--4.5, and 4.5--12 keV.}
\label{flc4E}
\end{center}
\vspace{-0.75 cm}
\end{figure}

In Fig.~\ref{flc4E} we show the folded light curve in four narrower energy bands (normalized to the average CR in each band). It proves that the pulse profile evolves from a single broad peak at the low-energy end to two narrow and very distinct  peaks at the high-energy end. The flux variability increases with energy since the CR varies by a factor $\simeq$ 2 at E $<$ 1 keV, but increases by a factor $\simeq$ 3.5 for E $>$ 4.5 keV. The same happens for the PF, which increases from 30 \% at E $<$ 1 keV up to above 60 \% at E $>$ 4.5 keV.

        \section{EPIC spectroscopy}
        \label{EPIC}

Since the light curve of \sxp\ shows that neither the flux nor the spectral properties of the source vary on the observation timescale, for each EPIC camera we accumulated its spectrum over the whole exposure. Each spectrum was extracted from the same region used for the light curve and was rebinned with a significance of at least 10 $\sigma$ for each energy bin. Then we used the \textsc{sas} tasks \textsc{rmfgen} and \textsc{arfgen} to generate the response matrices and ancillary files, respectively. We performed the spectral analysis in the energy range 0.2--12 keV, using version 12.9.1 of \textsc{xspec}, and calculated the spectral uncertainties and upper limits at the 90 \% confidence level for one interesting parameter. Since \sxp\ is in the SMC, we assumed the value of 62 kpc \citep{Graczyk+14} for the source distance. We  simultaneously fitted the spectra of the three EPIC cameras since their separate fits provided consistent results. Moreover, to account for possible uncertainties in instrumental responses, the relative normalizations of the three cameras were allowed to vary. For the spectral fitting we used the absorption model \textsc{tbnew} in \textsc{xspec}. To this end, we considered the interstellar abundances provided by \citet{WilmsAllenMcCray00} and the photoelectric absorption cross sections of \citet{Verner+96}. We described the interstellar absorption with two different components:  a Galactic component with solar abundances and column density fixed to $N_{\rm H}^{\rm GAL} = 6 \times 10^{20}$ cm$^{-2}$, the total Galactic absorption in the SMC direction \citep{DickeyLockman90}, and  an additional component ($N_{\rm H}^{\rm SMC+local}$), which accounts for the insterstellar absorption within the SMC and that local to the source, with free column density and abundances for elements heavier than helium set to 0.2, the canonical abundance value for the SMC \citep{RusselDopita92}.

\begin{figure}
\includegraphics[height=8.5cm,angle=-90]{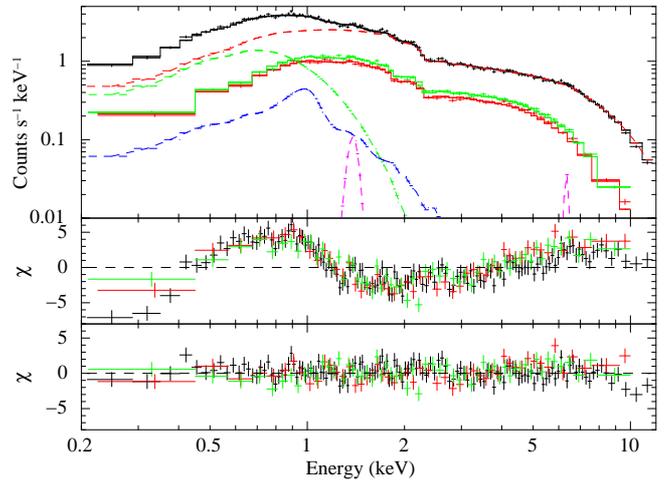}
\caption{Time-averaged spectrum of \sxp. Black, red, and green symbols  respectively represent \pn, MOS1, and MOS2 data. Upper panel: Superposition of the EPIC spectra with (for the \pn\ spectrum only) the best-fitting model composed of \textsc{pl+bb+apec} (red, green, and blue dashed lines, respectively) and the Gaussian components (magenta dashed lines). Middle panel: Data-model residuals obtained when fitting the spectra with a simple \textsc{pl} model. Lower panel: Data-model residuals obtained with the best-fitting model.}\label{epic_spectrum}
\end{figure}

We found that no single-component model was able to describe  the source spectrum successfully. For example, the fit with an absorbed PL model resulted in a best-fitting solution with \chisqnu/d.o.f. = 2.16/1214. In the medium panel of Fig.~\ref{epic_spectrum} we report the corresponding data-model residuals, which show the presence of a significant soft excess (SE) below E $\simeq$ 1 keV. We described this feature with a black-body (BB)  component and either a broad Gaussian emission line at $\simeq$ 0.95 keV or a model representing the spectral emission due to a collisionally ionized gas (\textsc{apec} in \textsc{xspec}). Moreover, in both cases we found two narrow emission lines at $\simeq$ 1.38 and 6.33 keV, which we modelled with two Gaussian emission lines of null width.

In Table~\ref{epic_fit} we list the best-fitting parameters obtained for these spectral models. Two different possibilities were considered for the metal abundance of the APEC component. On the one hand, it was fixed  at the estimated metallicity for the SMC \citep[$Z = 0.2 Z_\odot$,][]{RusselDopita92}, thus obtaining a fit quality equivalent to the value obtained with the BB+Gaussian model (\chisqnu\ = 1.05). On the other hand, we left the metal abundance free to vary and obtained an almost equivalent fit quality (\chisqnu\ = 1.06); in this case the best-fitting value of the abundance is 0.23$^{+0.31}_{-0.15}$, which is consistent with the expected value for the SMC. In all cases the best-fitting absorption value is $N_{\rm H}^{\rm SMC} = 1.2 \times 10^{21}$ cm$^{-2}$, which points out a significant absorption within the SMC or locally at the source.

The BB properties  in the three models are similar. The radius of this component is $\simeq$ 110 km and it contributes  $\simeq$ 3.5 \% to the total flux. On the other hand, the contribution of the APEC model is $\simeq$ 1.5 \%. The broad Gaussian emission line at $\simeq$ 0.95 keV of the first model can be attributed to either a mixture of several L$\alpha$ emission lines from Fe in various ionizations states (from {\sc Fe xviii} to Fe {\sc xx}) or a radiative recombination continuum (RRC) from O {\sc viii} - Ne {\sc ix}. The narrow Gaussian components at $\simeq$ 1.35 and 6.32 keV can be identified with Mg {\sc xi} and neutral Fe K$\alpha$ emission lines, respectively. However, regarding the Fe line we note that its energy is slightly lower than the typical value of 6.4 keV, and moreover that it is characterized by a small instrinsic width.

\begin{table*}
\caption{Results of the fit of the time-averaged EPIC spectra. The double-component continuum consists of a power-law and a black-body model, while the broad feature at 0.9--1 keV is described with either a broad Gaussian line or an APEC model. Both a fixed and a free value of the metal abundance are considered in the second case. In addition, two narrow Gaussian lines are needed to account for the positive residuals in the spectrum.}\label{epic_fit}
\vspace{-0.5 cm}
\begin{center}
\begin{tabular}{lccc} \hline
Model                                                           & \textsc{pl + bb + 3 Gaussian}      & \textsc{pl + bb + apec + 2 Gaussian}  & \textsc{pl + bb + apec + 2 Gaussian}       \\
parameter                                                       &                               & (fixed abundance)                       & (free abundance)                      \\ \hline
$N_{\rm H}^{\rm SMC+local}$ (10$^{21}$ cm$^{-2}$)                       & 1.16$\pm$0.13                   & 1.23$^{+0.09}_{-0.14}$                & 1.22$^{+0.21}_{-0.17}$          \\
$\Gamma$                                                        & 0.78$^{+0.01}_{-0.02}$        & 0.76$\pm$0.01                           & 0.76$\pm$0.02                         \\
Flux$_{\rm PL}$ (0.2-12 keV, $\times 10^{-11}$ \flux)           & 7.79$^{+0.06}_{-0.03}$         & 7.78$\pm$0.06                         & 7.78$\pm$0.06                         \\
$kT_{\rm BB}$ (eV)                                              & 177$^{+11}_{-13}$             & 171$^{+11}_{-14}$                       & 171$^{+14}_{-18}$                     \\
$R_{\rm BB}$ (km)                                               & 107$^{+18}_{-15}$             & 110$^{+25}_{-15}$                       & 111$^{+26}_{-15}$                     \\
Flux$_{\rm BB}$ (0.2-12 keV, $\times 10^{-12}$ \flux)           & 3.0$\pm$0.3                   & 2.8$\pm$0.3                             & 2.8$^{+0.4}_{-0.5}$                   \\
$kT_{\rm APEC}$ (keV)                                           & -                             & 1.09$^{+0.16}_{-0.09}$          & 1.09$^{+0.17}_{-0.09}$                \\
Abundance (\textsc{apec})                                       & -                             & 0.2 (fixed)                             & 0.23$^{+0.31}_{-0.15}$                \\
$N_{\rm APEC}$ ($\times 10^{-3}$ cm$^{-5}$)                     & -                             & 1.0$^{+0.6}_{-0.4}$                     & 0.9$^{+2.0}_{-0.8}$                   \\
Flux$_{\rm APEC}$ (0.2-12 keV, $\times 10^{-13}$ \flux)         & -                             & 9$^{+8}_{-2}$                           & 8$^{+22}_{-6}$                        \\
\multicolumn{4}{c}{Emission lines}      \\
$E_{\rm line1}$ (keV)                                           & 0.97$\pm$0.03                 & -                                       & -                                     \\
$\sigma_{\rm line1}$ (keV)                                      & 0.09$^{+0.05}_{-0.04}$        & -                                       & -                                     \\
Flux$_{\rm line1}$ ($\times 10^{-5}$ \fph)                      & 8$^{+5}_{-6}$                 & -                                       & -                                     \\
EW$_{\rm line1}$ (eV)                                           & 20$^{+11}_{-9}$               & -                                       & -                                     \\
$E_{\rm line2}$ (keV)                                           & 1.38$^{+0.03}_{-0.02}$        & 1.38$^{+0.03}_{-0.02}$          & 1.38$^{+0.03}_{-0.02}$                \\
$\sigma_{\rm line2}$ (keV)                                      & $<$ 0.04                      & 0 (fixed)                               & 0 (fixed)                             \\
Flux$_{\rm line2}$ ($\times 10^{-5}$ \fph)                      & 1.6$^{+0.7}_{-1.0}$           & 1.2$^{+0.8}_{-0.7}$                     & 1.3$^{+0.7}_{-0.8}$                   \\
EW$_{\rm line2}$ (eV)                                           & 6$^{+4}_{-3}$                 & 5$^{+4}_{-3}$                           & 5$\pm$3                               \\
$E_{\rm line3}$ (keV)                                           & 6.32$^{+0.09}_{-0.27}$        & 6.33$^{+0.09}_{-0.15}$          & 6.33$^{+0.09}_{-0.27}$                \\
$\sigma_{\rm line3}$ (keV)                                      & 0 (fixed)                     & 0 (fixed)                               & 0 (fixed)                             \\
Flux$_{\rm line3}$ ($\times 10^{-5}$ \fph)                      & 1.3$\pm$0.7                   & 1.2$\pm$0.7                             & 1.2$^{+0.8}_{-0.7}$                   \\
EW$_{\rm line3}$ (eV)                                           & 18$\pm$11                     & 18$\pm$11                               & 18$\pm$9                              \\
Flux$_{\rm BB}$/Flux$_{\rm TOT}$ (0.01-12 keV)                  & 3.8 \%                        & 3.5 \%                          & 3.6 \%                                \\
Flux$_{\rm APEC}$/Flux$_{\rm TOT}$ (0.01-12 keV)                & -                             & 1.4 \%                          & 1.3 \%                                \\
Unabsorbed flux (0.2-12 keV, $\times 10^{-11}$ \flux)           & 8.12$^{+0.06}_{-0.07}$        & 8.16$^{+0.08}_{-0.07}$          & 8.15$^{+0.11}_{-0.08}$                \\
Luminosity (0.2-12 keV, $\times 10^{37}$ \lum)                  & 3.53$^{+0.02}_{-0.03}$        & 3.54$^{+0.04}_{-0.03}$          & 3.54$^{+0.05}_{-0.03}$                \\
\chisqnu/d.o.f.                                                 & 1.05/1204                     & 1.05/1206                               & 1.06/1205                             \\ \hline
\end{tabular}
\end{center}
\end{table*}

\section{RGS Spectroscopy}
\label{rgs}

We extracted both the first-order and the second-order spectra from the data of both RGS instruments. Then, we used the {\sc sas} task {\sc rgscombine} to combine the two spectra of the same order. However, we verified that the count statistics of the combined second-order spectrum was very limited compared to those of the first-order, thus providing no additional information. Therefore, we considered only the first-order spectrum. We rebinned it with a minimum of 30 counts per bin and analysed it with {\sc xspec} in the energy range 0.4--2.1 keV.

\begin{figure}
\centering
\resizebox{\hsize}{!}{\includegraphics[angle=-90,clip=true]{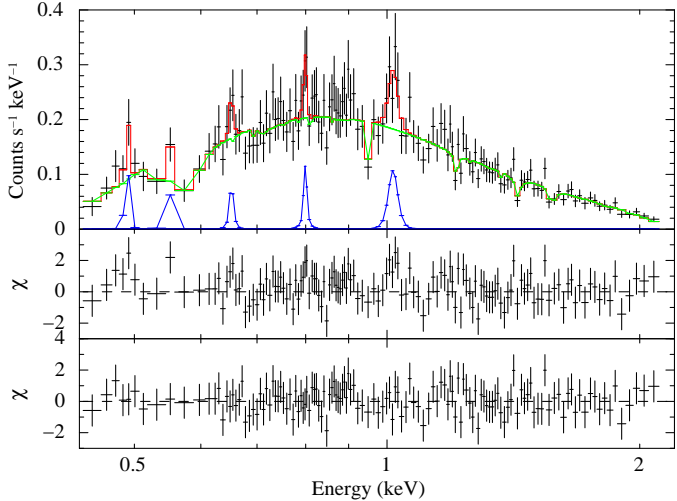}}
\caption{Combined spectrum for the first-order data of RGS1 and RGS2. Upper panel: Superposition of the spectrum with the best-fitting model composed by an absorbed power-law (green line) and additional Gaussian components (blues lines, Table~\ref{RGS_parameters}). Middle panel: Data-model residuals obtained when fitting the spectrum with a simple power-law model. Lower panel: Data-model residuals obtained with the best-fitting model (power-law plus Gaussian components).}
\label{RGS_spectrum}
\end{figure}

We started the spectral analysis by fitting the spectrum with a simple absorbed PL model. For the interstellar absorption we adopted the same approach used for the EPIC spectra, considering both a fixed Galactic component $N_{\rm H}^{\rm GAL} = 6 \times 10^{20}$ cm$^{-2}$ with solar abundances and a free component $N_{\rm H}^{\rm SMC+local}$ with SMC abundances. In this way we obtained a best-fitting solution which left emission residuals at energies E $\simeq$ 0.49, 0.55, 0.65, 0.80, and 1.02 keV (Fig.~\ref{RGS_spectrum}, middle panel). They were described with Gaussian components, whose parameters are listed in Table~\ref{RGS_parameters}. The intrinsic width of the lines at 0.65, 0.80, and 1.02 keV is well determined, while it is unconstrained for the two lines at 0.49 and 0.55 keV. Therefore, in the spectral fit  of these two lines we fixed the width value at 0. In this way we obtained a good description of the RGS spectrum (Fig.~\ref{RGS_spectrum}, lower panel). The normalization of the lines at 0.49, 0.80, and 1.02 keV is significant at the 99 \% confidence level, and then we can confirm their detection. On the other hand, for the other two lines at 0.55 and 0.65 keV the normalization is significant at the 90 \% confidence level only,  and hence their detection is only tentative. In Table~\ref{RGS_parameters} we propose a possible identification of each line. The features at 0.65 and 1.02 keV can be attributed to the Ly$\alpha$ lines of the H-like O {\sc viii} and Ne {\sc x}, respectively. The lines at 0.49 and 0.55 keV are marginally consistent with N {\sc vii} and O {\sc vii} lines, respectively. On the other hand, the identification of the feature at 0.80 keV is more uncertain, although it can be due to various Fe ionization stages.

\begin{table*}
\caption{Best-fit parameters of the emission lines identified in the RGS spectrum of \sxp.}\label{RGS_parameters}
\vspace{-0.5 cm}
\begin{center}
\begin{tabular}{cccccc} \hline
Observed                        & Ion                                           & Laboratory      & $\sigma$              & Flux                                  & EW                      \\
Energy                          &                                                       & Energy          &                               &                                               &                               \\
(eV)                            &                                                       & (eV)                    & (eV)                  &  (10$^{-5}$ \fph)             & (eV)                    \\ \hline
490$^{+5}_{-1}$         & N {\sc vii} (?)                       & 500.3                 & 0 (fixed)               & 9.3$^{+6.8}_{-5.4}$   & 9.7$^{+5.9}_{-8.4}$   \\
550$^{+12}_{-31}$       & O {\sc vii} (?)                       & 561.0                 & 0 (fixed)               & 9.0$^{+7.9}_{-6.7}$   & 12.8$^{+4.9}_{-11.5}$ \\
653$^{+4}_{-5}$         & O {\sc viii}                          & 653.5                 & 3.0$^{+9.4}_{-3.0}$     & 3.7$\pm$2.8   & 5.3$^{+3.8}_{-3.3}$   \\
799$^{+4}_{-3}$         & Fe {\sc xvii-xviii} (?)       & 793.0                 & 2.9$^{+8.4}_{-2.5}$     & 3.5$^{+1.9}_{-2.1}$   & 6.5$^{+3.7}_{-4.5}$   \\
1017$^{+8}_{-9}$        & Ne {\sc x}                            & 1022.0                & 11.0$^{+9.6}_{-4.4}$    & 8.0$^{+3.3}_{-4.0}$   & 20.3$^{+8.9}_{-8.5}$  \\ \hline
\end{tabular}
\end{center}
\end{table*}

As a final test, we also tried to describe the RGS spectrum with the best-fitting PL+BB+APEC model used for the EPIC spectra (section~\ref{EPIC}). The results we found in this way are fully consistent with the previous values, since the continuum component of the RGS spectrum can also be described  with the model used for the EPIC continuum. Moreover, in this case the residuals of the RGS spectrum are comparable to those obtained with a simple PL model (middle panel of Fig.~\ref{RGS_spectrum}), thus proving that it is necessary to describe the observed emission lines with specific Gaussian components.

\section{Phase-resolved spectral analysis}
\label{resolved_spectral_analysis}

The pulse profile of \sxp\ shows a clear energy dependence (Fig.~\ref{flc2E}), which suggests a spectral evolution with the spin phase. Therefore, we performed a phase-resolved spectral analysis by selecting two different spectra for each EPIC camera: spectrum A in the phase range $\Delta\Phi$ = 0.21--0.71 and spectrum B in the phase ranges $\Delta\Phi$ = 0--0.21, 0.71--1. In this way, we considered separately the parts of the folded light curve characterized by HR above or below 1, respectively.

We wanted to compare these two spectra with the best-fitting model used for the time-averaged spectrum in order to verify whether it can also provide an appropriate description  for these two spectra. Moreover, in this case we wanted to assess if the best-fitting parameters have consistent or completely different values. To this end, the two spectra were independently fit using the PL+BB+APEC model (with the metal abundance fixed at 0.2) for the spectral continuum. In Table~\ref{2spectra} we list the results obtained in this way.

\begin{table}
\caption{Best-fit parameters of the EPIC spectra A and B, assuming the PL+BB+APEC description of the spectral continuum and the metal abundance of the APEC component fixed at the estimated metallicity for the SMC ($Z = 0.2 Z_\odot$).}\label{2spectra}
\vspace{-0.5 cm}
\begin{center}
\begin{tabular}{cccc} \hline
Parameter                                       & Spectrum A                    & Spectrum B                      \\ \hline
$N_{\rm H}^{\rm SMC+local}$ (10$^{21}$ cm$^{-2}$)                       & 1.3$\pm$0.2                     & 1.2$^{+0.2}_{-0.3}$           \\
$\Gamma$                                        & 0.78$^{+0.01}_{-0.02}$        & 0.72$\pm$0.03                   \\
Flux$_{\rm PL}^{(a)}$                           & 10.7$\pm$0.1                  & 4.86$\pm$0.07                   \\
$kT_{\rm BB}$ (eV)                              & 180$^{+20}_{-40}$             & 165$\pm$14                      \\
$R_{\rm BB}$ (km)                               & 90$^{+50}_{-30}$              & 130$^{+30}_{-20}$               \\
Flux$_{\rm BB}^{(b)}$                           & 1.8$\pm$0.5                   & 3.4$^{+0.4}_{-0.3}$             \\
$kT_{\rm APEC}$ (keV)                           & 1.0$\pm$0.1                   & 1.2$^{+0.3}_{-0.2}$             \\
$N_{\rm APEC}$ ($\times 10^{-3}$ cm$^{-5}$)     & 1.4$^{+0.7}_{-0.6}$           & 1.0$^{+0.7}_{-0.6}$             \\
Flux$_{\rm APEC}^{(b)}$                         & 1.2$^{+0.6}_{-0.7}$           & 0.9$^{+0.7}_{-0.5}$             \\ \hline
\multicolumn{3}{c}{Emission lines}              \\
$E_{\rm line1}$ (keV)                           & 1.40$\pm$0.03                 & 1.37$\pm$0.04                   \\
$\sigma_{\rm line1}$ (keV)                      & 0 (fixed)                     & 0 (fixed)                       \\
Flux$_{\rm line1}^{(c)}$                        & 2.0$^{+1.2}_{-1.3}$           & 0.8$^{+0.9}_{-0.8}$             \\
EW$_{\rm line1}$ (eV)                           & 6$^{+4}_{-3}$                 & 5$\pm$5                 \\
$E_{\rm line2}$ (keV)                           & 6.3$^{+0.1}_{-0.5}$           & 6.3 (fixed)                     \\
$\sigma_{\rm line2}$ (keV)                      & 0 (fixed)                     & 0 (fixed)                       \\
Flux$_{\rm line2}^{(c)}$                        & 1.3$^{+0.6}_{-1.2}$           & 1.3$^{+0.8}_{-0.9}$             \\
EW$_{\rm line2}$ (eV)                           & 16$^{+12}_{-14}$              & 29$\pm$15                       \\ \hline
Flux$_{\rm BB}$/Flux$_{\rm TOT}$ (0.01--12 keV)  & 1.7 \%                        & 6.8 \%                  \\
Flux$_{\rm APEC}$/Flux$_{\rm TOT}$ (0.01--12 keV)& 1.5 \%                        & 2.2 \%                  \\
Unabsorbed flux$^{(d)}$                         & 1.10$\pm$0.01                 & 0.53$\pm$0.01                   \\
Luminosity$^{(e)}$                              & 4.78$\pm$0.05                 & 2.30$\pm$0.04                   \\
\chisqnu/d.o.f.                                 & 1.03/883                      & 1.14/469                        \\ \hline
\end{tabular}
\end{center}
$^{(a)}$ 0.2--12 keV, $\times 10^{-11}$ \flux
\\
$^{(b)}$ 0.2--12 keV, $\times 10^{-12}$ \flux
\\
$^{(c)}$ $\times 10^{-5}$ \fph
\\
$^{(d)}$ 0.2--12 keV, $\times 10^{-10}$ \flux
\\
$^{(e)}$ 0.2--12 keV, $\times 10^{37}$ \lum
\end{table}

We found that the SMC+local absorption is comparable in the two spectra and consistent with the best-fit value of the time-averaged spectrum. Therefore, the data show no evidence of absorption variability. The total source flux decreases by a factor of $\simeq$ 2 between spectrum A and B. This reduction is essentially due to the PL component, which in both cases contributes  more than 90 \% to the total flux. Its photon index is comparable to the value of the time-averaged spectrum in the case of spectrum A, and only slightly lower for spectrum B. Both the BB and the APEC components are significant at the  99 \% confidence level in both spectra, but their behaviour is very different: on the one hand, the APEC flux is almost constant, so that its relative contribution to the total flux slightly increases in  spectrum B; on the other hand, the BB flux doubles, so that its relative contribution to the total flux increases by a factor of $\simeq$ 4. We note that the BB temperature  between the two phases is very similar and fully consistent with the time-averaged value. Therefore, the flux variation of this component is mainly due to the different size of its emission region. Instead the APEC temperature is slightly higher in spectrum B. The two emission lines detected in the time-averaged spectrum are much less evident in these two phase-resolved spectra. In spectrum A both lines are significant only at the  90 \% confidence level, while in spectrum B the line at 1.38 keV remains undetected. Moreover, the energy of the Fe line is unconstrained in spectrum B. Therefore, in this case we had to fix it at the best-fit value of the time-averaged spectrum. In this way, we found that the Fe line is significant at the  90 \% confidence level. This result is very probably due to the lower count statistics of these two spectra compared to the time-averaged spectrum.

We also performed  a simultaneous fit of the two spectra in order to further constrain the variability of the continuum components. In this case we assumed a common value for \nh\ and fixed the energy and width of the emission lines since the count statistics are not high enough to constrain them. For the fit we first considered two different solutions for the PL+BB+Gaussian model of the continuum: (1) a common value of the PL photon index and BB temperature, leaving both normalizations free to vary, and (2) a common BB component. We found that in the first case the null hypothesis probability (NHP) of the best-fit model is $\simeq$ 0.03, while in the second  it reduces to $\simeq 9 \times 10^{-5}$, even if the photon index $\Gamma$ of the PL component is left free to vary between the two spectra. This means that a spectral fit with a constant BB component is statistically unacceptable and can be rejected. We repeated the same type of test also with the PL+BB+APEC model of the continuum, with three different solutions: (1) a common value of the PL photon index and of the BB and APEC temperatures and independent normalizations, (2) a common BB component, and (3) a common APEC component. For the best-fit model of solution (1) we found that NHP $\simeq$ 0.03. For solution (2) the spectral fit is significantly worse, since NHP $< 2 \times 10^{-4}$ even if both $\Gamma$ and the APEC temperature are left independent. In case (3), instead, the spectral fit is only slightly worse (NHP = 6 $\times 10^{-3}$);  moreover, it becomes comparable to solution (1) if the two BB temperatures are independent. These results show that while there is no evidence of variability of the APEC component, a constant BB component is clearly rejected by the data.

In summary, the spectral variability observed between the two phase ranges can be attributed to the variation in  the relative contribution of the continuum components since the BB flux increases and the PL flux decreases between phase ranges A and B, while the APEC remains almost constant. This behaviour explains the energy-dependent pulse shape shown in Fig.~\ref{flc2E}.

         \section{\Swift/XRT and \OGLE\ observations}

\subsection{\Swift/XRT}
\label{xrt}

The X-ray telescope (XRT) on board the \Swift\ satellite performed a dozen  snapshot observations of \sxp\ between 2017 April 7 and May 1, with exposure times between a few hundred seconds and $\sim$ 3.3 ks each. In all cases the XRT instrument was operated in Windowed Timing (WT) mode. The data of all the observations were reduced following the standard procedures\footnote{http://www.swift.ac.uk/analysis/xrt/index.php}, extracting source and background events from circular regions of radius 20 pixels. The event arrival times were reported to the solar system barycentre by using the tool \textsc{barycorr}. Because of the unknown orbital solution of SXP59.0, we did not correct the data for the orbital motion.

\begin{figure}
\begin{center}
\includegraphics[width=6cm,angle=-90]{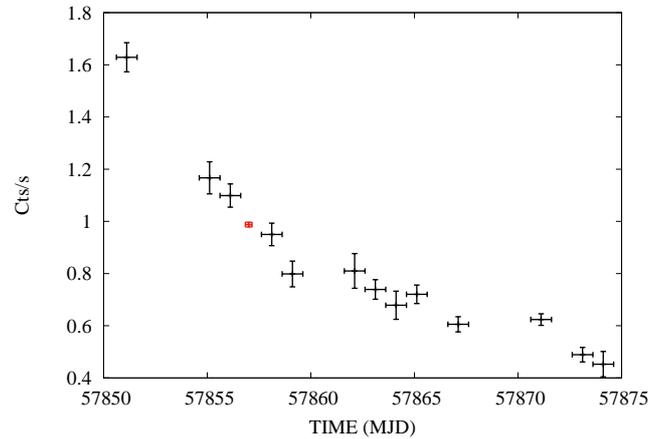}
\caption{Flux evolution of \sxp\ during the 2017 outburst, as observed with \Swift/XRT. The red symbol corresponds to the epoch and flux (with its uncertainty) of the \XMM\ observation.}
\label{swift_lc}
\end{center}
\vspace{-0.75 cm}
\end{figure}

\begin{figure}
\begin{center}
\includegraphics[width=6cm,angle=-90]{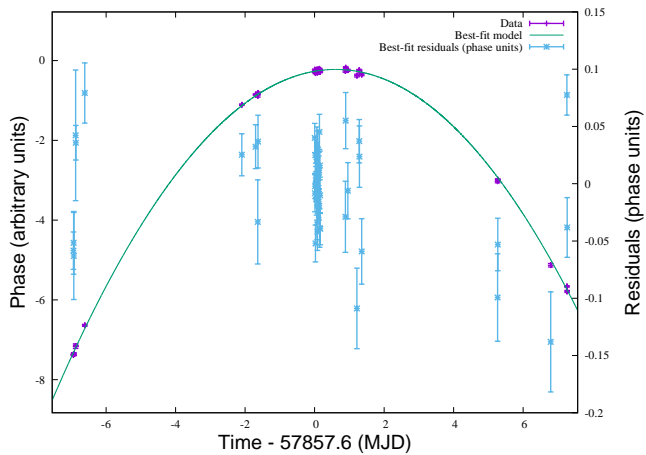}
\caption{Results of the phase-connection analysis of the pulsations of \sxp.}
\label{phase_connection}
\end{center}
\end{figure}

The \Swift/XRT lightcurve is shown in Fig.~\ref{swift_lc}. The red symbol represents the \Swift/XRT count rate estimated from the source flux measured with \XMM\ (CR$_{\rm XRT} = f_{\rm XMM}$/(8.1$\times 10^{-11}$ \flux) cts s$^{-1}$). The \Swift/XRT followed the outburst decay which lasted at least 25 days. Based on the ephemerides provided by \citet{Galache+08}, we estimated the epoch of the centroid of the two maxima of the 122 d X-ray modulation which preceded and followed the time period covered by the \Swift\ observations: they are MJD 57801 $\pm$ 21 and 57923 $\pm$ 21, respectively. Therefore, the \Swift\ and \XMM\ observations were performed far from the expected epoch of flux peak. In each \Swift/XRT observation, we detected the source pulsation at $\simeq$ 58.9 s. Thanks to the tight temporal distance between the \XMM\ and \Swift/XRT observations, we were able to  perform a phase-connection of the pulsations. Starting from the EPIC \pn\ data, we phase-connected with a linear function the \Swift/XRT observation taken $\sim$1 day after. However, the linear function was not able to fully account for the phase variation and therefore we added a parabolic correction ($\dot{\nu}$), which significantly improved the fit. We then phase-connected all the other remaining observations until the orbital effects (not taken into account) started to be important and did not allow us to continue the analysis (Fig.~\ref{phase_connection}). In particular, we were able to connect $\simeq$ 14 days and found this solution: \ps\ = 58.95392161(2) s, \pdot\ = -1.160(3)$\times 10^{-7}$ s s$^{-1}$, and $\chi^2_{\nu}=3.9$ for 48 dof. The reference epoch is MJD 57957.6212479.

\subsection{\OGLE}
\label{ogle}

\begin{figure}
\begin{center}
\includegraphics[width=6.5cm,angle=-90]{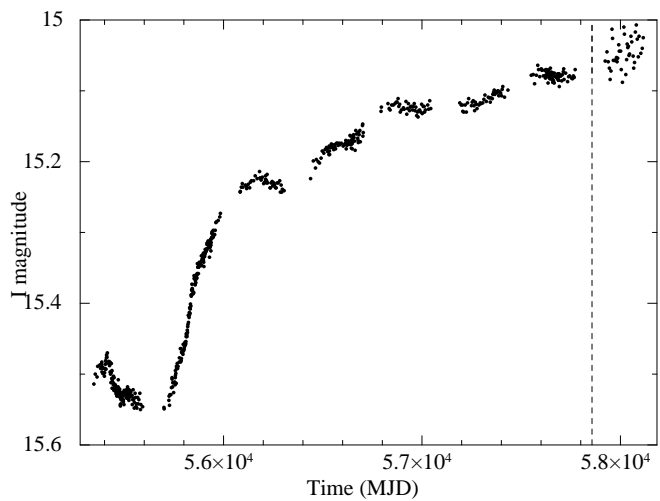}
\caption{Long-term light curve of the optical counterpart of \sxp\ based on I-magnitude OGLE IV data. The dashed vertical line at MJD 57857 corresponds at the epoch of the \XMM\ observation.}
\label{ogle_lc}
\end{center}
\end{figure}

In order to put the \XMM\ observation of \sxp\ in the context of the evolution of its optical counterpart, we used the OGLE monitoring system of the X-ray variables (XROM\footnote{http://ogle.astrouw.edu.pl/ogle4/xrom/xrom.html}, \citealt{Udalski08}) to look for the OGLE IV \citep{Udalski+15} optical light curve. In Fig.~\ref{ogle_lc} we show the long-term OGLE IV light curve, which includes data from May 2010  to December 2017. It shows that starting from MJD 55720 (June 2011) the source brightness has continuously increased until the latest observations. The dashed vertical line at MJD 57857 marks the epoch of the \XMM\ observation. It was performed near the end of the OGLE IV monitoring campaign, but within a period which was not covered by any OGLE observation.

\section{Discussion}

\begin{figure}
\begin{center}
\includegraphics[width=6.5cm,angle=-90]{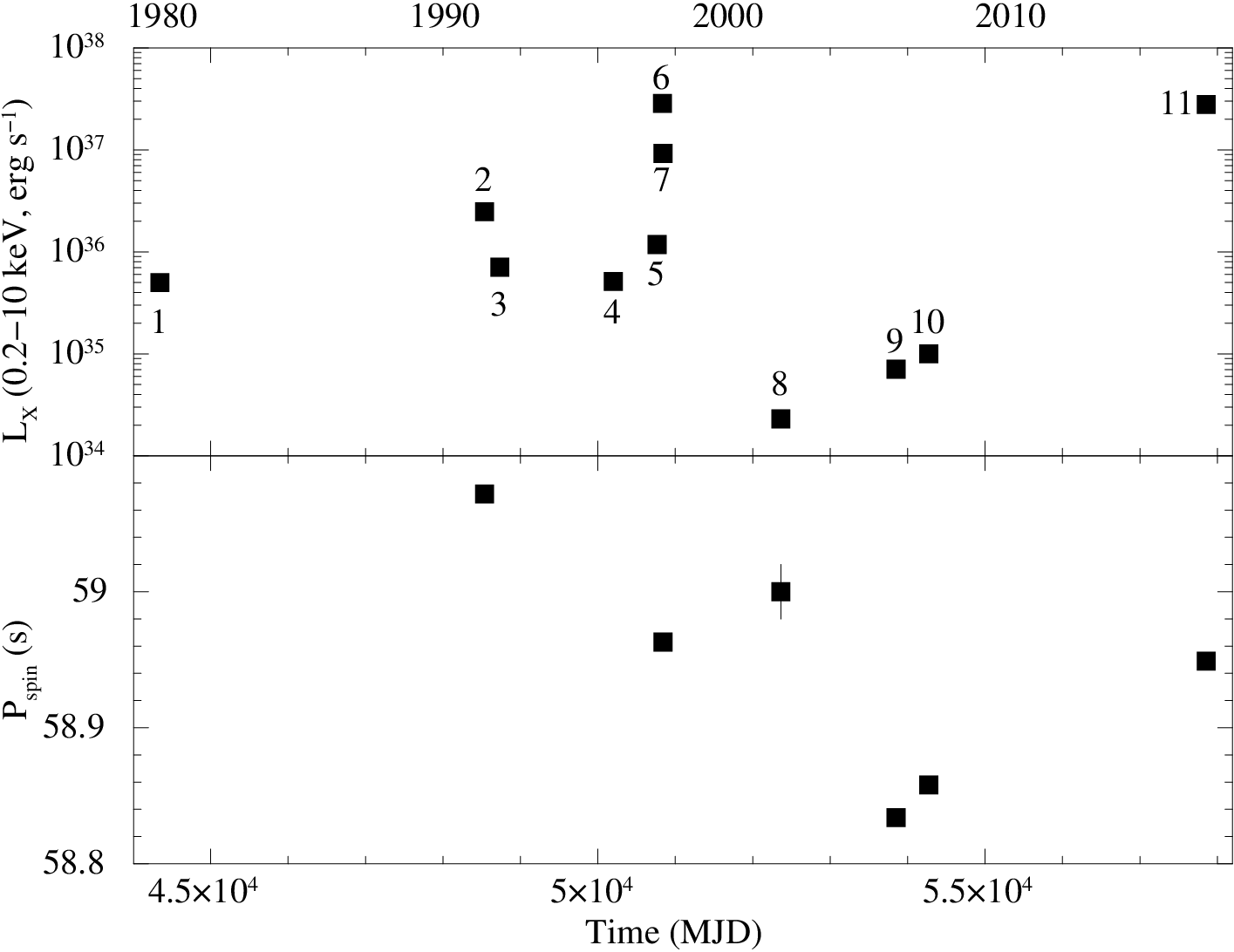}
\caption{Long-term evolution of the 0.2--12 keV luminosity (\textit{upper panel}) and of the spin period (\textit{lower panel}) of \sxp. Date and instrument: (1) \Einstein/PSPC: 15-04-1980 \citep{Bruhweiler+87}; (2) \ROSAT/PSPC: 08-10-1991 \citep{Israel+98}; (3) \ROSAT/PSPC: 22-04-1992 \citep{Israel+98}; (4) \ROSAT/HRI: 26-04-1996 \citep{Israel+98}; (5) \ASCA: 14-11-1997 \citep{Yokogawa+00}; (6) \RXTE: 22-01-1998 \citep{Marshall+98}; (7) \SAX: 28-01-1998 \citep{Santangelo+98}; (8) \XMM: 30-03-2002 \citep{Sasaki+03}; (9) \Chandra: 26-04-2006 \citep{Laycock+10}; (10) \XMM: 23-06-2007 \citep{Haberl+08}; (11) \XMM: 14-04-2017 (this work)}
\label{overall_lc}
\end{center}
\end{figure}

Since its discovery with \Einstein\ in 1980, \sxp\ has been observed several times at very different luminosity levels. In the upper panel of Fig.~\ref{overall_lc} we report the long-term evolution of its 0.2--10 keV luminosity, as estimated based on the observed flux level and assuming a source distance of 62 kpc. It shows that the source luminosity varies by more than three orders of magnitude between $\sim 10^{34}$ and $\sim 10^{37}$ \lum. For the first two decades the source was detected only during rather bright states, with \lx\ \gsim\ $10^{36}$ \lum. In particular, at the beginning of 1998 (MJD 50835) the source luminosity increased of over 1.5 orders of magnitude in $\sim$ two months, up to \lx\ $\simeq 3 \times 10^{37}$ \lum. Afterwards, the high-throughput X-ray telescopes \XMM\ and \Chandra\ detected it as well at significantly lower flux levels,  \lx\ \lsim\ $10^{35}$ \lum. The \XMM\ observation performed in April 2017 caught \sxp\ at one of the highest flux levels ever detected, possibly comparable only to that observed with \RXTE\ in 1998 (assuming that the latter does not include any contribution of other nearby sources). This ToO observation allowed us to perform a detailed timing and spectral analyses of this source down to $\simeq$ 0.2 keV.

The lower panel of Fig.~\ref{overall_lc}, where we give all the published period values obtained in previous observations, shows that the period varies in the range 58.8--59.1 s. Moreover, according to \citet{Klus+14}, all the pulse periods measured with \RXTE\ between 1998 and 2011 are in the range $\simeq$ 58.3--59.3 s. This period variability cannot be due to the orbit of the neutron star (NS) around the companion star. If we assume a mass $M \sim 10 \msole$ for the Be star, the orbital period $P_{\rm orb} \simeq$ 122 d implies an average orbital velocity $v_{\rm orb} \simeq$ 90 km/s. The corresponding period variation is $\Delta P = P \times v_{\rm orb}/c \simeq$ 0.02 s, a value much smaller than the observed variation in period values. Therefore, is it very likely that the period variability is due to the pulsar spin-up and spin-down  during the outburst and quiescence phases, respectively. The pulse period measured with \XMM\ in 2017 is well within the range of values previously reported. This implies that, although the source can have experienced various outburst and quiescence phases since 2011, its average spin period has not changed significantly. The phase-connection of the pulsations reported in Section~\ref{xrt} shows that the pulsar undergoes a very high spin-up during the outburst phases. Therefore, it is very likely that these spin-ups are balanced by a significant slow-down during the quiescent phases, when the pulsar is most probably in the propeller state.

In the last years  three additional transient Be binary pulsars in the SMC have been observed at high spectral resolution during an outburst: \rxj\ \citep{Sidoli+15}, \smc\ \citep{LaPalombara+16}, and \IGR\ \citep{LaPalombara+18}. Compared to \sxp, these sources are characterized by a shorter pulse period and, in the first two cases, by a higher outburst luminosity. However, a direct comparison of these binary pulsars with \sxp\ is useful, since they have similar spectral and timing properties. To this end, in Table~\ref{transients} we report the main properties of these four sources.

\begin{table*}
\caption{Comparison of the main timing and spectral parameters of the transient BeXRBs \rxj\ \citep{Sidoli+15}, \smc\ \citep{LaPalombara+16}, \IGR\ \citep{LaPalombara+18}, and \sxp\ (this work).}\label{transients}
\vspace{-0.5 cm}
\begin{center}
\begin{tabular}{cccccccc} \hline
Parameter                                       & \multicolumn{2}{c}{\rxj$^{(a)}$}                              & \multicolumn{2}{c}{\smc$^{(a)}$}                                & \multicolumn{2}{c}{\IGR$^{(a)}$} & \sxp  \\ \hline
\lx\ (0.2-12 keV, $\times 10^{37}$ \lum)                & \multicolumn{2}{c}{7}                         & \multicolumn{2}{c}{14}                          & \multicolumn{2}{c}{3.6}       & 3.5     \\
\po\ (days)                                     & \multicolumn{2}{c}{82}                        & \multicolumn{2}{c}{18.4}                                & \multicolumn{2}{c}{35.6}      & 122.1 \\
$P_{\rm spin}$ (s)                      & \multicolumn{2}{c}{2.76}                      & \multicolumn{2}{c}{2.37}                                & \multicolumn{2}{c}{11.58}     & 58.95 \\
PF (\%)                                         & \multicolumn{2}{c}{8.9}                       & \multicolumn{2}{c}{35}                          & \multicolumn{2}{c}{43}        & 53      \\
\nh\ (10$^{20}$ cm$^{-2}$)      & \multicolumn{2}{c}{2.3$^{+0.6}_{-0.5}$}               & \multicolumn{2}{c}{18$\pm$3}            & \multicolumn{2}{c}{1.0$^{+0.1}_{-0.2}$}       & 12$\pm$1        \\
$kT_{\rm BB}$ (eV)                      & 93$\pm$5      & -                     & 135$^{+14}_{-11}$       & -             & 218$^{+13}_{-14}$     & -             & 171$^{+11}_{-14}$ \\
$R_{\rm BB}$ (km)                       & 350$^{+80}_{-50}$     & -     & 320$^{+125}_{-95}$      & -     & 50$^{+6}_{-5}$        & -             & 110$^{+25}_{-15}$       \\
f$_{\rm BB}$/f$_{\rm PL}$ (\%)    & 1.7   & -                     & 3.1     & -                                     & 1.6   & -                                     & 3.5     \\
$kT_{\rm APEC}$ (keV)           & -     & 0.21$\pm$0.03 & -     & 1.22$^{+0.07}_{-0.10}$        & - & 1.13$^{+0.10}_{-0.08}$      & 1.09$^{+0.16}_{-0.09}$        \\
$N_{\rm APEC}$ ($\times 10^{-3}$ cm$^{-5}$)             & -             & 25$^{+8}_{-6}$  & -     & 5$\pm$1       & -                             & 4$\pm$1                         & 1.0$^{+0.6}_{-0.4}$\\
f$_{\rm APEC}$/f$_{\rm PL}$ (\%)  & -     & 7     & -                             & 1.8     & -                                     & 4.5                                   & 1.4 \\
N {\sc vii}                             & \multicolumn{2}{c}{yes}                               & \multicolumn{2}{c}{yes}                         & \multicolumn{2}{c}{yes}       & yes     \\
O {\sc vii}                             & \multicolumn{2}{c}{no}                                & \multicolumn{2}{c}{yes}                         & \multicolumn{2}{c}{no}        & yes     \\
O {\sc viii}                    & \multicolumn{2}{c}{yes}                               & \multicolumn{2}{c}{yes}                         & \multicolumn{2}{c}{yes}       & yes     \\
Ne {\sc ix}                             & \multicolumn{2}{c}{yes}                               & \multicolumn{2}{c}{yes}                         & \multicolumn{2}{c}{yes}       & no      \\
Ne {\sc x}                              & \multicolumn{2}{c}{no}                                & \multicolumn{2}{c}{yes}                         & \multicolumn{2}{c}{no}        & yes     \\
Mg {\sc xi}                             & \multicolumn{2}{c}{no}                                & \multicolumn{2}{c}{no}                          & \multicolumn{2}{c}{yes}       & yes     \\
Si {\sc xiii}                           & \multicolumn{2}{c}{no}                                & \multicolumn{2}{c}{yes}                         & \multicolumn{2}{c}{no}        & no      \\
Si {\sc xiv}                            & \multicolumn{2}{c}{no}                                & \multicolumn{2}{c}{yes}                         & \multicolumn{2}{c}{no}        & no      \\
E$_{\rm Fe K\alpha}$ (keV)      & \multicolumn{2}{c}{6.6}                               & \multicolumn{2}{c}{6.6}                         & \multicolumn{2}{c}{6.4}       & 6.3     \\
$d_{\rm BB}$ (km)                       & \multicolumn{2}{c}{3000}                              & \multicolumn{2}{c}{1800}                                & \multicolumn{2}{c}{400}       & 570     \\
$R_{\rm m}$ (km)                        & \multicolumn{2}{c}{900}                               & \multicolumn{2}{c}{740}                         & \multicolumn{2}{c}{1100}      & 1100    \\ \hline
\end{tabular}
\end{center}
$^{(a)}$ For this source the APEC component is considered an alternative to the BB component.
\end{table*}

The source pulsations of \sxp\ were detected, for the first time, even at E $<$ 1 keV. The pulse profile observed with \XMM\ is clearly energy dependent: while it shows a single broad maximum at low energies, two distinct narrow peaks arise at high energies. The pulsed fraction is $>$ 40 \% at E $<$ 2 keV and  increases further with the energy. Moreover, the spectral hardness increases with the source flux. Similar results were obtained during the previous observations of the source outbursts in 1991 with \ROSAT\ \citep{Israel+98} and in 1998 with \SAX\ \citep{Santangelo+98}. In both cases the pulse profile was asymmetric and the pulsed fraction was above 40 \%. Moreover, \SAX\ data showed two distinct peaks above 1.5 keV. On the other hand, the source observations at low luminosity level performed with \XMM\ in 2002 \citep{Sasaki+03} and with \Chandra\ in 2006 \citep{Laycock+10} revealed a single-peak sinusoidal profile, with an almost constant hardness ratio. Our results for \sxp\ are also similar  to what we observed for \rxj, \smc, and \IGR. All of them showed a double-peaked profile, with a high pulsed fraction in the case of \smc\ and \IGR. Since the source was observed at a rather high luminosity level, it is very probable that the double-peaked profile is due to a fan-beam emission geometry. 

The average EPIC spectrum of \sxp\ is dominated by a rather hard absorbed PL component, with photon index $\Gamma \simeq$ 0.8 and total absorption due not only to the Galactic absorption in the SMC direction but also to a local component in the SMC itself. Both results are in agreement with the results provided by the previous observations performed with \ROSAT\ \citep{Israel+98}, \SAX\ \citep{Santangelo+98}, \XMM\ \citep{Sasaki+03}, and \Chandra\ \citep{Laycock+10}. In all cases the best-fit photon index was in the range $\Gamma \sim$ 0.8--1 and the total absorption was \nh\ $\sim$ (1--3)$\times10^{21}$ cm$^{-2}$, thus significantly higher than the Galactic value in the SMC direction ($N_{\rm H}^{\rm GAL} = 6 \times 10^{20}$ cm$^{-2}$, \citealt{DickeyLockman90}). However, the spectral fit with this single-component model leaves large residuals at energies below $\sim$ 1 keV, which reveal the presence of a previously unknown soft excess above the main PL component. We fitted this soft excess with the sum of a soft (kT $\sim$ 0.2 keV) BB and either a broad Gaussian component at 0.95 keV or a hot (kT $\sim$ 1 keV) thermal plasma model (APEC). In both cases the sum of the two soft components contributes  $\simeq$ 5 \%\ to the total source  luminosity. It is interesting to note that the properties of both thermal components are very similar to those of the corresponding components observed in \rxj, \smc, and \IGR\ (Table~\ref{transients}). In particular, this is true in the case of \IGR, which was observed at a luminosity level almost equal to that of \sxp. However, we emphasize that, in the case of \sxp, both components are necessary to obtain an acceptable spectral fit, while either the BB or the APEC was required for the other sources. This is an important difference since it implies that the observed soft excess of \sxp\ is due to the coexistence of two different emission processes.

The \XMM\ observation of \sxp\ during its 2017 outburst allowed us to detect previously unknown narrow emission features in the source spectrum. The high-resolution RGS spectrum showed the presence of emission lines due to N, O, and Ne, and of some other emission features, which could be attributed to L-shell lines from Fe at various ionization levels. The EPIC spectrum revealed two emission lines at 1.38 and 6.32 keV, which we attributed to K$\alpha$ emission from Mg and neutral Fe, respectively (although they are characterized by a negligible intrinsic width and, in the case of the Fe line, the best-fit energy is slightly lower than the typical value of 6.4 keV). Similar emission features were already observed in the spectra of \rxj, \smc, and \IGR\ during their outbursts (Table~\ref{transients}). This is particularly true in the case of \IGR, since it is the only other source which showed a K$\alpha$ line from neutral Fe (while the other two sources showed an emission line consistent with ionized Fe). From this point of view, it is interesting to note that both sources were observed at the same luminosity level.

The phase-resolved spectral analysis of the EPIC data allowed us to study the dependence of the pulse profile as a function of the energy and to investigate the spectral variability along the pulse. To this end, we analyzed separately the spectra corresponding to, respectively, the high  hard state and the low  soft state, which are chacterized by a flux difference of $\simeq$ 50 \%. We found that, in both cases, all  three continuum components of the time-averaged spectrum (PL, BB, and APEC) were required to obtain an acceptable fit, and that  the PL  always contributed  more than 90 \% to the total flux. Moreover, the PL photon index and the temperatures of the two thermal components were always consistent with those of the averaged spectrum. However, the behaviour of the single components was significantly different. While the APEC flux remained almost constant and the PL halved its flux, the BB doubled it. These findings were confermed by the simultaneous analysis of the two spectra which showed that a constant BB component is inconsistent with the data.

The \XMM\ observation of \sxp\ was performed during an outburst, when the source luminosity increased  to \lx\ $\simeq 3.5 \times 10^{37}$ \lum. This luminosity level implies a high accretion rate onto the NS, which very likely occurs through an accretion disc. This scenario is supported by the high spin-up rate measured during the outburst. In this case, according to \citet{Hickox+04}, the observed SE can be due to the combination of two different types of processes: emission from photoionized or collisionally heated diffuse gas and reprocessing of hard X-rays from the NS by optically thick accreting material, most probably at the inner edge of the accretion disc. We have found that the SE of \sxp\ is due to the sum of a BB and a APEC component, where only the BB is clearly pulsating. If this component is due  to reprocessing of the primary radiation, and $\Omega$ is the solid angle subtended by the reprocessing region, the relation between the reprocessed and the primary luminosity is given by $L_{\rm BB}$ = ($\Omega$/4$\pi$) $L_{\rm X}$. On the other hand, $L_{\rm BB} = \Omega d_{\rm BB}^2 \sigma T^4_{\rm BB}$, where $d_{\rm BB}$ is the distance of the reprocessing site from the source of the primary emission. This implies that $d_{\rm BB}^2 = L_{\rm X}$/($4\pi \sigma T^4_{\rm BB}$). Since for \sxp\ $L_{\rm X} \simeq 3.5 \times 10^{37}$ \lum\ and $T_{\rm BB} \simeq$ 0.17 keV, the estimated distance of the BB component from the central NS is $d_{\rm BB} \simeq 570$ km. This value is the estimated radius of the inner edge of the accretion disc, which should be comparable to the magnetospheric radius of the accreting NS: $R_{\rm m} \sim 1.5 \times 10^8 m^{1/7} R_6^{10/7} L_{37}^{-2/7} B_{12}^{4/7}$ cm, where $m = M_{\rm NS}$/$\msole$, $R_6 = R_{\rm NS}$/(10$^6$ cm), $B_{12} = B_{\rm NS}$/(10$^{12}$ G), and $L_{37} = L_{\rm X}$/(10$^{37}$ \lum) \citep{Davies&Pringle81}. If we assume $m$ = 1.4, $R_6$ = 1, and $B_{12}$ = 1, since for \sxp\ $L_{37}$ = 3.5, we obtain $R_{\rm m} \simeq$ 1100 km. This value is comparable to $d_{\rm BB}$ within a factor $\simeq$ 2, thus supporting the hypothesis that the observed BB component arises from the reprocessing of the primary radiation at the inner edge of the accretion disc. Moreover, this value is equal to that obtained for \IGR\ and comparable to those obtained for \rxj\ and \smc\ (Table~\ref{transients}). For the estimate of the magnetospheric radius we have assumed the same values of the NS parameters (mass, radius, and magnetic field) for all the sources. Therefore, the different values of $R_{\rm m}$ are only related to the different luminosity levels: it decreases as the luminosity increases. For all the binary pulsars listed in Table~\ref{transients}, $d_{\rm BB}$ is comparable with $R_{\rm m}$ within a factor of 2--3. This difference can have various physical or geometrical causes, since the NS magnetic field can be different by $10^{12}$ G; moreover, our estimate of $R_{\rm m}$ can be affected by a tilted and/or warped accretion disc.

Regarding the non-pulsating APEC component, from its normalization of $\simeq 10^{-3}$ cm$^{-5}$ we can derive a value of $\simeq 5.5 \times 10^{59}$ cm$^{-3}$ for the emission measure $n^2 V$. Assuming a gas density $n < 10^{12}$ cm$^{-3}$, as we did in the case of the other sources listed in Table~\ref{transients}, and a spherical emitting region for the optically thin plasma, we obtain a radius of $R \gsim 5 \times 10^{11}$ cm. Therefore, the size of this component is much larger than the inner radius of the accretion disc. We note that, assuming $v_{\rm w} \simeq$ 100 km s$^{-1}$ for the wind velocity of the Be star and $v_{\rm r}^2 = v_{\rm w}^2 + v_{\rm orb}^2$ for the relative velocity between the wind and the NS, from the relation $R_{\rm acc}$ = 2GM/$v_{\rm r}^2$ we obtain $R_{\rm acc} \simeq 2 \times 10^{12}$ cm for the accretion radius around the NS. This value is consistent with the estimated size of the APEC component. Therefore, it is possible that this component is related to the shock region around the NS caused by the accreted wind from the companion star.

The timing and spectral properties of the BB component support its identification with the reprocessed primary radiation from the inner edge of the accretion disc. As proposed in the case of \IGR\ \citep{LaPalombara+18}, at each phase along the pulse period, the inner edge of the disc is swept by the beamed primary emission. Only a limited section of the disc surface is illuminated by the beam. Moreover, it is possible that the disc geometry allows us to see only a limited fraction of the disc edge. For these reasons, the reprocessed component can be detected only when this visible portion of the disc is hit by the primary beam. It is possible that  the Fe line observed in the time-averaged spectrum also has the same origin, as we have demonstrated in the case of \IGR, but for \sxp\ we cannot prove it because we lack  any evidence of its variability. On the other hand, the other lines observed in the EPIC and RGS spectra are probably due to photoionized matter in regions above the disc since they cannot be described with the BB+APEC model.

\section{Conclusions}

The \XMM\ observation of \sxp\ reported in this paper completes our programme of ToO observations of X-ray binary pulsars in the SMC. Thanks to the low interstellar absorption towards the SMC direction and the high flux level reached by these sources during their outbursts, these observations allowed us to perform detailed spectral and timing analysis and to study the physical processes that occur in these sources.

In all cases we detected the observed source at \lx $\gsim 4 \times 10^{37}$ \lum\ and obtained high-resolution spectra with high count statistics. We obtained consistent results for these pulsars:
\begin{itemize}
\item the pulsed emission can be observed over the whole energy range between 0.2 and 12 keV, but the pulse profile is energy dependent and the pulsed fraction increases with energy;
\item the continuum spectrum is dominated by a hard ($\Gamma < 1$) PL component, but shows a soft excess at E $\lsim$ 1 keV, which contributes a small percentage of the total flux and is mainly due to a variable BB component;
\item the EPIC spectrum is characterized by an emission feature which can be identified with a neutral or ionized Fe K$\alpha$ transition line, while the RGS spectrum shows low-energy features due to ionized N, O, and Ne;
\item the BB component which dominates the SE and the Fe K$\alpha$ emission line can be attributed to the reprocessing of the primary radiation by the optically thick material at the inner edge of the accretion disc, while the other emission features are most probably due to photoionized plasma above the accretion disc; only in the case of \sxp\ did we find evidence for a collisionally heated thermal plasma, probably due to the shocked wind accreted from the companion Be star.
\end{itemize}

\begin{acknowledgements}
We acknowledge the use of public data from the \Swift\ data archive. We acknowledge the financial contribution from the agreement ASI-INAF I/037/12/0. PE acknowledges funding in the framework of NWO Vidi award A.2320.0076 and in the framework of the project ULTraS, ASI-INAF contract N.\,2017-14-H.0.
\end{acknowledgements}

\bibliographystyle{aa}
\bibliography{biblio}

\end{document}